\documentclass[conference]{IEEEtran}
\IEEEoverridecommandlockouts
\usepackage{cite}
\usepackage{textcomp}

\usepackage[T1]{fontenc}
\usepackage{lmodern}
\usepackage{color}
\usepackage[pdftex]{graphicx}
\usepackage{booktabs}
\usepackage{multirow}
\usepackage{bm}
\usepackage{bbm}
\usepackage{url}
\usepackage{amsmath,amsthm,amssymb,cases}
\usepackage{subfigure}
\usepackage{subfiles}
\usepackage{algorithm, algpseudocode}
\usepackage{mathtools}
\usepackage{mathrsfs}
\usepackage{diagbox}

\graphicspath{{./}}

\def\BibTeX{{\rm B\kern-.05em{\sc i\kern-.025em b}\kern-.08em
    T\kern-.1667em\lower.7ex\hbox{E}\kern-.125emX}}
\begin{document}

\title{Delayed Coding Scheme for Channels with Insertion, Deletion, and Substitution Errors}

\author{%
  \IEEEauthorblockN{Ryo Shibata and Hiroyuki Yashima}
  \IEEEauthorblockA{Dept. of Information and Computer Technology, Faculty of Engineering,\\Tokyo University of Science, Tokyo 125-8585, Japan\\
  Email: \{r\_shibata, yashima\}@rs.tus.ac.jp}
}

\maketitle
\begin{abstract}
We propose a new coding scheme, called the \textit{delayed coding (DC)} scheme, for channels with insertion, deletion, and substitution (IDS) errors.
The proposed scheme employs delayed encoding and non-iterative detection and decoding strategies to manage the transmission of multiple codewords in a linear code.
In the DC scheme, a channel input sequence consists of subblocks of multiple codewords from the previous to current time instances.
At the receiver side, the maximum \textit{a posteriori} detection applies to the received sequences that contain information of the codeword at the current time instance, where priorly decoded codewords aid the detection.
The channel code decoding is then performed, and extrinsic messages are exploited for the codeword estimations of the following time instances.
We show that the rate achievable with the DC scheme over the IDS channel approaches the symmetric information rate of the channel.
Moreover, we show the excellent asymptotic and finite-length performances of the DC scheme in conjunction with low-density parity-check codes.
\end{abstract}
\section{Introduction}
Recently, synchronization error correction codes have garnered much attention owing to their occurrence in future storage devices\cite{dwm, dna}.
An insertion or deletion of a symbol is the most well-known instance of synchronization errors.
To tackle random insertion and deletion (ID) errors, several constructions based on low-density parity-check (LDPC)\cite{ldpc1,ldpc2} codes have been proposed in the past two decades \cite{water,marker,scid,my_irr,my_irr2} and achieved promising results in approaching the symmetric information rate (SIR) of the channel.
Davey and Mackay \cite{water} and Wang et al. \cite{marker} studied the concatenated codes of outer LDPC codes with inner synchronization codes (e.g., marker and watermark codes).
The inner codes serve to discriminate positions of ID errors by exploiting their non-uniform input distributions.
Currently, standalone LDPC codes (i.e., without an inner code) are being investigated \cite{scid,my_irr,my_irr2}.
These standalone codes have synchronization recovery structures (e.g., structured irregularities \cite{scid} and low-degree check nodes\cite{my_irr}).
Both concatenated and standalone codes are used in conjunction with a maximum \textit{a posteriori} (MAP) detector, which infers \textit{a posteriori} probabilities (APPs), prior to the belief-propagation (BP) decoder of an LDPC code.
In general, the re-estimation of the APPs after each BP iteration (i.e., a joint iterative detection and decoding strategy) enhances decoding accuracy, at the expense of increased complexity and decreased throughput.
To the best of our knowledge, there are no coding schemes that approach the SIRs of ID-type channels without the joint iteration.

Another concern for most of the existing coding schemes is the fact that they do not possess universal performance behaviors for noisy channels with ID errors, such as ID-substitution (IDS) and ID-AWGN channels.
This means that different code structures are required to suppress different noise levels (or variance) \cite{marker,my_irr2}.
For instance, inner synchronization codes are utterly useless for correcting random noise errors as they simply result in a rate loss.

In this paper, we propose a new coding scheme, called the \textit{delayed coding (DC)} scheme, which considers the transmission of a large number of codewords in a linear code.
In the DC scheme, subblocks of multiple codewords from the previous to current time instances comprise a channel input sequence.
Such a structure serves to remove the memory effects of ID errors from a particular codeword.
At the receiver side, the codewords are successively decoded by employing a non-iterative detection and decoding algorithm.
In estimating a particular codeword, the MAP detection is applied to the received sequences containing information regarding that codeword, where priorly decoded codewords function as pilot markers and aid the detection.
We show that the rates achievable with the DC scheme over the IDS channel universally approach the SIRs, regardless of substitution probabilities.
Furthermore, we demonstrate that the DC scheme with LDPC codes approaches the SIRs in terms of the asymptotic and finite-length performances, with reduced complexity compared to the existing coding scheme.

The remainder of this paper is organized as follows.
In Section\ II, we introduce the proposed system model.
In Section\ III, we show the achievable information rates (AIRs) for the proposed DC scheme. In Section\ IV, we demonstrate the asymptotic and finite-length performances when using LDPC codes.
Finally, in Section\ V, we present our conclusions and future directions.
\section{System model}
Throughout this paper, upper case letters are used to denote random variables, whereas their realizations are denoted by lower case letters.
\subsection{IDS channel model}
We consider the concatenation of a binary input--output ID channel with a binary symmetric channel, i.e., an IDS channel \cite{marker, my_irr2}.
Let $\bm{X}=(X_{1},X_{2},\ldots, X_{n}) \in \{0,1\}^{n}$ denote the input sequence.
In this channel, each input bit $X_{t}$, $t \in \{1,\ldots,n\}$, is deleted with probability $p_{\rm id}/2$, replaced by two random bits (i.e., insertion) with probability $p_{\rm id}/2$, or transmitted with probability $1-p_{\rm id}$.
During transmission, $X_{t}$ is flipped with a probability of $p_{\rm s}$.
Through this process, $\bm{X}$ is mapped to the output sequence $\bm{Y}=(Y_{1},Y_{2},\ldots, Y_{n'}) \in \{0,1\}^{n'}$, where $n'$ can be larger or smaller than $n$.
For ID-type channels, the process $Y_{t}$ is typically formulated as the output of a hidden Markov model (HMM).
Similar to \cite{scid,my_irr,my_irr2}, we adopt the channel state process $S_{t}$ given by a simple random walk in the presence of reflecting boundaries over the set $\{-D_{\rm max},...,+D_{\rm max}\}$,
where $D_{\rm max}$ is the maximum allowed drift ($D_{\rm max}>0$).
The state $S_{t}=i$ represents the timing drift at time $t$ (the difference between the number of transmitted and received bits is $i$).
The maximum drift $D_{\rm max}=8$ is used in this study based on observations in \cite{my_irr}.

For ID-type channels, instead of capacity, the SIR is generally used as an achievable information rate.
The SIR is defined as
\begin{align*}
  C_{\rm SIR}  = \lim_{n \to \infty} \frac{1}{n} I(\bm{X}; \bm{Y})\mid_{p(\bm{X} = \bm{x})=2^{-n}},
\end{align*}
that is, the mutual information rates between the i.i.d.\ equiprobable inputs and corresponding outputs.
The estimation of $C_{\rm SIR}$ was carried out in \cite{simar,ar2} based on the Shannon--McMillan--Breiman theorem for discrete stationary ergodic random processes.
\subsection{DC scheme}
\begin{figure}[t]
\centering
\includegraphics[width=.97\linewidth]{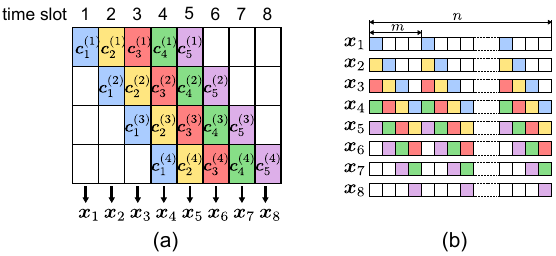}
\caption{Example illustrating the proposed DC encoding scheme with $\bm{T}=(0,1,2,3)$ and $L=5$: (a) block level and (b) bit level.}
\label{fig:dc_enc}
\end{figure}
We now describe the proposed DC scheme for channels with ID errors.
The main idea underlying our scheme is inspired by a synchronization marker \cite{marker} and recently proposed modulation technique called \textit{delayed bit interleaved coded modulation} \cite{dbcim}.

Let $\bm{c}_{t} = (c_{1},c_{2},\ldots,c_{n}) \in \{0,1\}^{n}$ represent the interleaved codeword selected from a binary linear code $\mathcal{C}$ at time $t \in \{1,\ldots,L\}$, where $L$ is the number of transmitted codewords.
Hereafter, no distinction is made between codewords and ``interleaved'' codewords.
At the transmitter side, each codeword is divided into $m$ subblocks $\bm{c}_{t}^{(1)},\bm{c}_{t}^{(2)},\ldots,\bm{c}_{t}^{(m)}$ of $n/m$ bits (for simplicity, $n$ is assumed to be divisible by $m$).
For $i \in \{1,\ldots,m\}$, each subblock $\bm{c}^{(i)}_{t}$ is associated with a delay $T_{i} \in \{0,\ldots, T_{\rm max}\}$, where $T_{\rm max}$ denotes the maximum allowed delay.
We call $\bm{T}=(T_{1}, T_{2}, \ldots, T_{m})$ a delay scheme.
While using a delay scheme $\bm{T}$, the $t$-th transmitted sequence $\bm{x}_{t} \in \{0,1\}^{n}$, $t \in \{1,\ldots, L+T_{\rm max}\}$, is obtained by interleaving $\bm{c}^{(1)}_{t-T_{1}}, \bm{c}^{(2)}_{t-T_{2}},\ldots, \bm{c}^{(m)}_{t-T_{m}}$ as follows:
\begin{align}
  \bm{x}_{t} &= \bm{c}^{(1)}_{t-T_{1}} || \bm{c}^{(2)}_{t-T_{2}} ||\cdots || \bm{c}^{(m)}_{t-T_{m}},\\
  &= (c^{(1)}_{t-T_{1},1}, c^{(2)}_{t-T_{2},1}, \ldots c^{(m)}_{t-T_{m}, 1}, c^{(1)}_{t-T_{1}, 2}, c^{(2)}_{t-T_{1}, 2}, \nonumber \\
  &\qquad \qquad \ldots, c^{(m)}_{t-T_{m}, 2}, c^{(1)}_{t-T_{1}, 3}, \ldots, c^{(m)}_{t-T_{m}, n/m}),
\end{align}
where $||$ denotes the interleaving operator.
The sequence $\bm{x}_{t}$ is transmitted over an IDS channel, and the corresponding output $\bm{y}_{t}$ is observed at the receiver.
Figure \ref{fig:dc_enc} shows the configuration of transmitted sequences for $\bm{T}=(0,1,2,3)$ and $L=5$, where $m=4$ and $T_{\rm max}=3$.
As shown in the figure, to start and terminate the encoding process, known bits (white-colored blocks in Fig.\ \ref{fig:dc_enc}) participate in some transmitted sequences.
In this study, we set these known-bits to be independent and uniformly distributed (i.u.d.), and the delay scheme $\bm{T} = (0,1,\ldots, T_{\rm max})$ is employed for a given $T_{\rm max}$.

As explained in the following subsection, the delay and known bits lead to a performance improvement in detection (re-synchronization).
However, they also incur a rate loss.
Given a code rate $R$ of a component code $\mathcal{C}$, the transmission rate of the DC scheme is
\begin{align}
  R_{\rm DC} = R \left( \frac{L}{L+T_{\rm max}}\right).
\end{align}
We can see that, for large $L$ $(L\gg T_{\rm max})$, $R_{\rm DC}$ approaches $R$.
Note that the DC scheme does not impose a constant rate-loss as required for the use of inner codes.

For future use, we introduce the sets $\mathcal{D}_{j}$ and $\mathcal{\tilde{D}}_{j}$ characterized by a given DC scheme.
For $j \in \{0,\ldots,T_{\rm max}\}$, $\mathcal{D}_{j} = \{i \mid T_{i}=j\}$ and  $\mathcal{\tilde{D}}_{j} = \{i \mid j < T_{i}\}$ denote the index sets of subblocks $i$ at delay equal to $j$ and larger than $j$, respectively.
\subsection{Chained detection and decoding algorithm}
\begin{figure}[t]
\centering
\includegraphics[width=.99\linewidth]{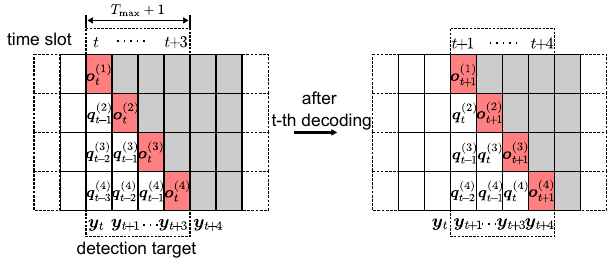}
\caption{Example illustrating chained detection with $\bm{T}=(0,1,2,3)$ at time instances $t$ and $t+1$.}
\label{fig:dc_dec}
\end{figure}
We explain the chained detection and decoding algorithm used in the DC scheme in this section.
This algorithm proceeds by successively estimating original codewords $\bm{c}_{t}$, $t \in \{1,\ldots, L\}$, one by one starting from $t=1$ to $L$, where priorly decoded codewords help to recover the current codeword in the detection.
The MAP detector, implemented with the Bahl--Cocke--Jelinek--Raviv (BCJR) algorithm, is employed here.

For $i \in \{1, \ldots, L\}$, $\bm{q}_{i}=(\bm{q}_{i}^{(1)}, \bm{q}_{i}^{(2)}, \ldots, \bm{q}_{i}^{(m)})$ denotes the extrinsic soft-messages output from the $i$-th code decoder, whereas $\bm{o}_{i} = (\bm{o}_{i}^{(1)}, \bm{o}_{i}^{(2)}, \ldots, \bm{o}_{i}^{(m)})$ denotes the sequence of log-APP ratios provided by the detector.
These $\bm{q}_{i}$ and $\bm{o}_{i}$ correspond to the $i$-th codeword $\bm{c}_{i}$ and are also expressed as using $m$ divided subblocks.
Let us now look at the procedure for estimating $\bm{c}_{t}$.
The $t$-th estimation can operate after the $(t-1)$-th estimation is completed and the sequences $\bm{y}_{t},\bm{y}_{t+1},\ldots,\bm{y}_{t+T_{\rm max}}$ are received.
First, for all $i \in \{0,\ldots, T_{\rm max}\}$, we run the MAP detection for the sequence $\bm{y}_{t+i}$ given priori messages $\bm{q}_{t-T_{j}}^{(j)}$, $j \in \mathcal{\tilde{D}}_{i}$, and exploit a part of the log-APP ratios sequence $\bm{o}_{t}^{(j)}$, $j \in \mathcal{D}_{i}$.
This is illustrated in Fig.\ \ref{fig:dc_dec}, where $\bm{T}=(0,1,2,3)$ (i.e., $\mathcal{D}_{i}=\{i + 1\}$ for $i\in \{0,\ldots,T_{\rm max}\}$, $\mathcal{\tilde{D}}_{0}=\{2,3,4\}$, $\mathcal{\tilde{D}}_{1}=\{3,4\}$, $\mathcal{\tilde{D}}_{2}=\{4\}$, and $\mathcal{\tilde{D}}_{3}=\phi$).
As shown in the figure, the log-APP ratios $\bm{o}_{t}$ are collected from different $(T_{\rm max}+1)$ received sequences and then fed to the $t$-th code decoder, such as the BP decoder of an LDPC code.
Finally, the decoder estimates $\bm{c}_{t}$, operating as it would on a memoryless channel, and sends the extrinsic messages $\bm{q}_{t}^{(j)}$, $j \in \mathcal{\tilde{D}}_{0}$ to subsequent estimations.

As shown in Fig.\ \ref{fig:dc_dec}, at the $t$-th estimation round, the MAP detector can use reliable \textit{a priori} information corresponding to the pre-estimated codewords $\bm{c}_{t-3}$, $\bm{c}_{t-2}$, and $\bm{c}_{t-1}$.
Moreover, the known bits provide perfect \textit{a priori} information to the detector.
In other words, these reliable bits behave as pilot symbols.
The chained estimation strategy benefits from this advantage, which prevents the DC scheme from using a joint iterative detection and decoding strategy.

Before leaving this section, we compare the computational complexity between the existing and DC schemes in terms of the MAP detection.
It goes without saying that the non-iterative scheme typically used with concatenated codes has the lowest complexity because it applies the MAP detection only once.
For the iterative scheme, the execution number of the MAP detection depends on the received sequences, channel parameters, code structures, and decoder configuration.
For instance, if we use a standalone LDPC code with a BP-based decoding algorithm, the detection is performed at every iteration until all parity check equations are satisfied, or a predefined number of iterations is reached.
Meanwhile, for the DC scheme, the MAP detection is performed just $(T_{\rm max}+1)$ times before channel code decoding.
Note that these multiple detections are parallelizable because they are performed on received sequences that are independent of each other.
Such a structure and non-iterative data flow strategy can improve the throughput and latency of the receiver and mitigate the effect of delay due to the DC scheme.
\section{Achievable information rates}
\begin{figure}[t]
\centering
\includegraphics[width=.99\linewidth]{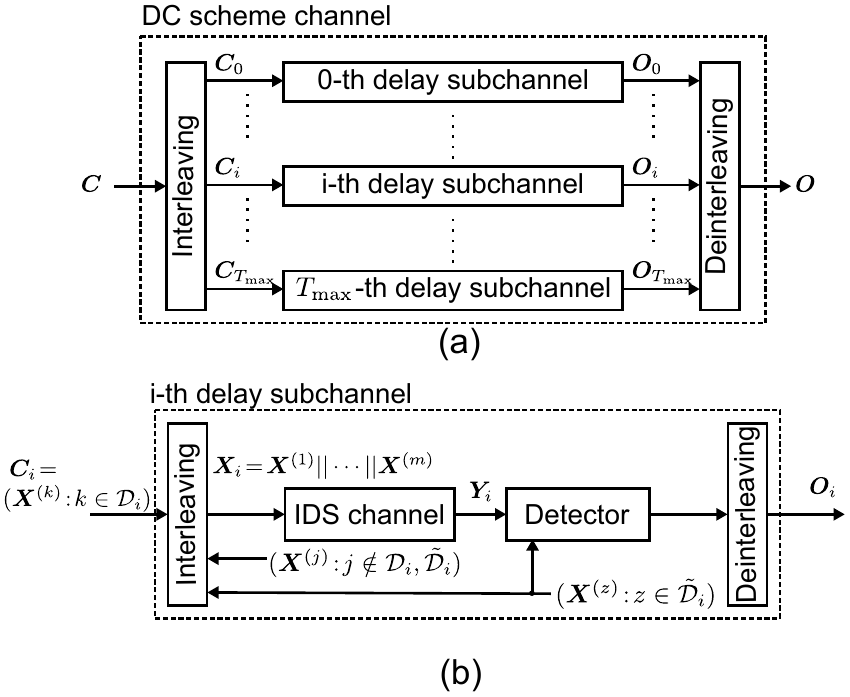}
\caption{(a) Delay scheme channel. (b) $i$-th delay subchannel.}
\label{fig:dc_ar}
\end{figure}
In this section, we present AIRs for the DC scheme.
As the BCJR algorithm has high complexity, strategies that apply the detection step only once are often considered for channels with states.
The AIR with such a strategy is called the \textit{BCJR-once rate} \cite{isi,once,dnaonce}, which is defined as the mutual information between the channel inputs and corresponding log-APP ratios.
To evaluate the AIRs of our scheme, we focus on the transmission of a particular random codeword.
Specifically, we consider the concatenation channel of a DC, an IDS channel, and MAP detection, whose input and output are $\bm{C}\in \{0,1\}^{n}$ and $\bm{O}\in \mathbb{R}^{n}$, respectively.
This channel, called a \textit{DC scheme channel (DCSC)}, is illustrated in Fig.\ \ref{fig:dc_ar} and can be decomposed into $(1+T_{\rm max})$ parallel independent channels, called \textit{delay subchannels}, related to each delay.

Let $\bm{C}_{i}$ and $\bm{O}_{i}$ denote parts of the interleaved $\bm{C}$ and $\bm{O}$ of lengths $n|\mathcal{D}_{i}|/m$ corresponding to the $i$-th delay subchannel, respectively.
We assume that the input sequence $\bm{C}$ is i.u.d.\, and define the BCJR-once rate over a DCSC as
\begin{align}
  R_{\rm DC} = \lim_{n \to \infty}\frac{1}{n}  \sum_{i=0}^{T_{\rm max}}\sum_{j=1}^{n|\mathcal{D}_{i}|/m} I(C_{i,j};O_{i,j}).
\end{align}
In the following, instead of estimating $R_{\rm DC}$ directly, we determine it by separately considering a BCJR-once rate for each delay subchannel.
In the $i$-th delay subchannel, the IDS channel input $\bm{X}_{i} = \bm{X}^{(1)}||\bm{X}^{(2)}||\cdots || \bm{X}^{(m)}$ consists of three parts: (i) bits corresponding to $\bm{C}_{i}$ (i.e., $\bm{X}^{(k)}, k \in \mathcal{D}_{i}$), (ii) priorly estimated bits (i.e., $\bm{X}^{(z)}, z\in \mathcal{\tilde{D}}_{i}$), and (iii) the other bits (i.e., $\bm{X}^{(j)}, j \notin \mathcal{D}_{i}, \mathcal{\tilde{D}}_{i}$).
Suppose that the codewords transmitted before $\bm{C}$ are successfully decoded, then the BCJR-once rate $R^{(i)}$ for the $i$-th channel is given by
\begin{align}
  R^{(i)} = \lim_{n_{\rm s} \to \infty} \frac{1}{n_{\rm s}|\mathcal{D}_{i}|} \sum_{j=1}^{n_{\rm s}|\mathcal{D}_{i}|} I(C_{i,j}; O_{i,j}),
\end{align}
where $n_{\rm s} := n/m$, and the MAP detector provides
\begin{align*}
  o_{i,j} = \ln \frac{\Pr\left[C_{i,j}=0 \mid \bm{Y}_{i}=\bm{y}_{i}, \{\bm{X}^{(z)} = \bm{x}^{(z)} ; z\in \mathcal{\tilde{D}}_{i}\} \right]}{\Pr\left[C_{i,j}=1 \mid \bm{Y}_{i}=\bm{y}_{i}, \{\bm{X}^{(z)} = \bm{x}^{(z)} ; z\in \mathcal{\tilde{D}}_{i}\}\right]},
\end{align*}
because the detector can exploit perfect information about bits for $\bm{X}^{(z)}$, $z \in \mathcal{\tilde{D}}_{i}$.
Ultimately, the BCJR-once rate over a DCSC is given by $R_{\rm DC} = \frac{1}{m}\sum_{i=0}^{T_{\rm max}}|\mathcal{D}_{i}| R^{(i)}$.
The rate $R^{(i)}$ can be numerically computed in a similar manner to \cite{isi}.
We note that, assuming a DCSC is a memoryless channel, any rate $r < R_{\rm DC}$ is achievable.
Although the channel in fact has memory (i.e., $\Pr(O_{i,j}, O_{i,z} \mid C_{i,j},C_{i,z}) \neq \Pr(O_{i,j}\mid C_{i,j}) \Pr(O_{i,z}\mid C_{i,z})$), the use of a delay scheme $\bm{T}$ with large $m$ and $|\mathcal{D}_{i}|=1$ for all $i \in \{0,\ldots, T_{\rm max}\}$ makes a DCSC virtually memoryless.
\begin{figure}[t]
\centering
\includegraphics[width=.95\linewidth]{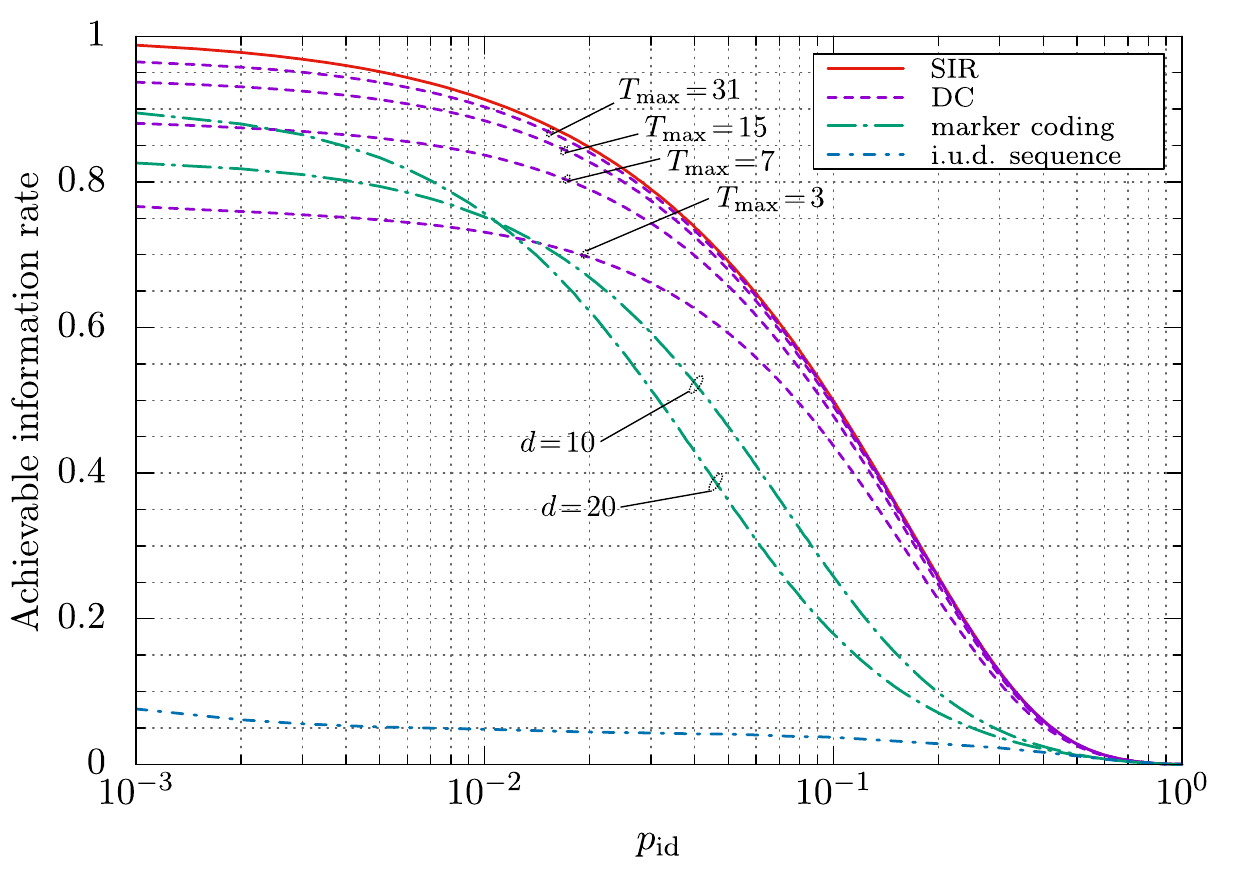}
\caption{BCJR-once rates for the DC scheme, marker coding scheme, and i.u.d.\ inputs, on the ID channel ($p_{\rm s}=0$).}
\label{fig:res1}
\end{figure}
\begin{figure}[t]
\centering
\includegraphics[width=.95\linewidth]{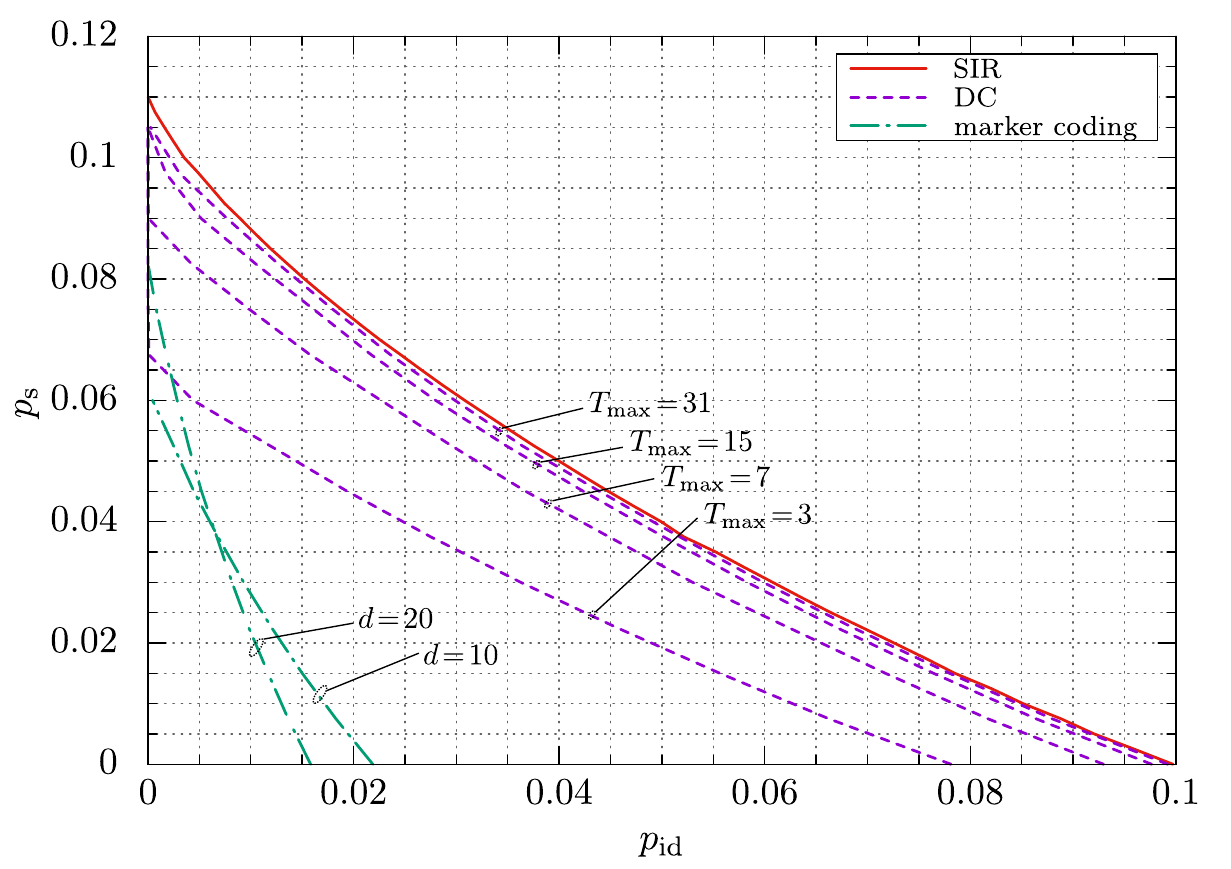}
\caption{BCJR-once limits for the DC scheme and marker coding scheme.}
\label{fig:res2}
\end{figure}

Figure \ref{fig:res1} shows the BCJR-once rates $R_{\rm DC}$ with $T_{\rm max}\in \{3,7,15,31\}$ over ID channels (i.e., $p_{\rm s}=0$).
It also shows three other types of curves: (1) the SIR $C_{\rm SIR}$ as computed using the Arnold--Loeliger method\cite{simar}, (2) the BCJR-once rate for the marker coding scheme\cite{marker} with two-bit periodic markers inserted into the i.u.d.\ sequence at every $d\in \{10,20\}$ bits, and (3) the BCJR-once rate with i.u.d.\ inputs (i.e., the rate achievable with a standalone linear code using MAP detection applied only once).
As shown in the figure, the BCJR-once rate of the DC scheme outperforms the marker coding scheme and i.u.d.\ inputs for a wide range of ID probabilities.
For $T_{\rm max}=15$ and $31$, the gap between the once rate and SIR is almost closed without depending on ID probabilities $p_{\rm id}$.
Figure \ref{fig:res2} shows the pairs $(p_{\rm id}, p_{\rm s})$ such that the SIR/BCJR-once rate is equal to $0.5$ for IDS channels, as a function of $p_{\rm s}$ (we call the pairs the SIR/BCJR-once limits).
It is clear that the DC scheme also performs well over the IDS channel, and it universally approaches the SIR when $T_{\rm max}=15$ and $31$ regardless of $p_{\rm s}$.
Such a universal property does not apply to the coding schemes with inner codes (e.g., watermark\cite{marker} and marker\cite{marker} codes) because they themselves incur a rate loss for correcting random errors.
\section{Performance of DC scheme with LDPC codes}
This section presents the asymptotic and finite-length performances of the DC scheme with LDPC codes.
\subsection{Asymptotic performance analysis}
For memoryless symmetric channels, density evolution\cite{ldpc2} is well known as a useful technique to predict the asymptotic performance of LDPC codes with BP-based decoding algorithms.
This technique is also extended to channels with states or/and asymmetric nature\cite{scid,isi,once}, where LDPC coset codes are employed.
Building on their results, we can establish the density evolution for the DC scheme with LDPC coset codes.
When using LDPC codes, the chained estimation algorithm of the DC scheme involves $(T_{\rm max}+1)$ MAP detection and subsequent BP decoding for each codeword/time instant $t\in \{1,\ldots,L\}$, starting from $t=1$.
Here, our interest is the convergence behavior of the BP decoding for the $t$-th codeword.
We assume that the $t'$-th ($t' < t$) BP decoder produces a probability of error that converges to zero as the number of BP iterations tends to infinity, i.e., the prior decoding round is successful.
Under this assumption, the transmission of the $t$-th codeword is reduced to the transmission over a DCSC.
Therefore, by running density evolution for a DCSC, we can predict the asymptotic threshold (i.e., the BP threshold) of an ID probability $p^{\ast}_{\rm id}$ over the IDS channel with a given $p_{\rm s}$.

In this study, we primarily investigate the irregular LDPC code ensembles introduced in \cite{ldpc2}.
They are specified by degree distribution pairs $\lambda(x)=\sum^{d_{\rm v}}_{i=2} \lambda_{i}x^{i-1}$ and $\rho(x)=\sum^{d_{\rm c}}_{i=2} \rho_{i}x^{i-1}$, where $\lambda_i$ (resp.\ $\rho_i$) denotes the fraction of the edges connected to the degree $i \geq 2$ variable (resp.\ check) nodes and $d_{\rm v}$ (resp.\ $d_{\rm c}$) denotes the maximum degree of the variable (resp.\ check) nodes.
The design rate of an LDPC code is given by $R=1- (\int_{0}^{1} \rho(x) dx/\int_{0}^{1} \lambda(x) dx)$.

Let us now compare the BP thresholds of some existing irregular LDPC codes of a rate $R=0.5$, whose degree distributions are shown in Table \ref{table:degree}.
The ID and IDS codes are designed for the iterative scheme over the ID ($p_{\rm s}=0$) and IDS ($p_{\rm s}=0.04$) channels, respectively.
In Table \ref{table:thr}, we show the BP thresholds of these codes with the DC scheme.
For comparison, we also show the SIRs and BP threshold of the $C(J=3, K=6, L)$ SC-LDPC code \cite{sc} (variable node degree $J$, check node degree $K$, and coupling length $L$).
We can see that, with the iterative scheme, the ID, IDS, and SC-LDPC codes exhibit outstanding performance, whereas the Bi-AWGN code cannot correct IDS errors.
However, interestingly, for the Bi-AWGN code with the DC scheme, the BP thresholds approach the corresponding BCJR-once rates (see Fig.\ \ref{fig:res1}) given $T_{\rm max}$ and are close to the SIRs, without depending on $p_{\rm s}$.
This is also observed for the DC scheme with the SC-LDPC code.
These results indicate that the DC scheme provides its universal performance even in conjunction with practical linear codes and decoding schemes.
For this to be realized, careful code selection is necessary.
More precisely, it seems reasonable to use codes designed for random errors, instead of ID and IDS codes, which have many low-degree check nodes.
\begin{table}[t]
\centering
\caption{Degree distributions of the Bi-AWGN \cite{ldpc2}, ID \cite{my_irr}, and IDS codes \cite{my_irr2}.}
\label{table:degree}
\scalebox{0.84}{
\begin{tabular}{|c|c|c||c|c|c||c|c|c|}
\hline
\multicolumn{3}{|c||}{Bi-AWGN\cite{ldpc2}} & \multicolumn{3}{c||}{ID\cite{my_irr}} & \multicolumn{3}{c|}{IDS\cite{my_irr2}} \\ \hline \hline
$i$   & $\lambda_{i}$   & $\rho_{i}$  & $i$  & $\lambda_{i}$  & $\rho_{i}$ & $i$   & $\lambda_{i}$  & $\rho_{i}$  \\ \hline
2     & 0.24426         &             & 2    & 0.39884        & 0.10018    & 2     & 0.24136        & 0.08363     \\
3     & 0.25907         &             & 3    & 0.45664        &            & 3     & 0.49091        &             \\
4     & 0.01054         &             & 4    &                & 0.30375    & 5     & 0.11696        &             \\
5     & 0.05510         &             & 9    &                & 0.22102    & 7     & 0.07771        & 0.45892     \\
7     &                 & 0.25475     & 12   & 0.14452        & 0.37505    & 9     & 0.00506        & 0.13615     \\
8     & 0.01455         & 0.73438     &      &                &            & 11    &                & 0.30544     \\
9     &                 & 0.01087     &      &                &            & 12    & 0.06800        & 0.01586     \\
10    & 0.01275         &             &      &                &            &       &                &             \\
12    & 0.40373         &             &      &                &            &       &                &             \\ \hline
\end{tabular}
}
\end{table}
\begin{table}[t]
\centering
\caption{Comparison of BP thresholds $p_{\rm id}^{\ast}$ with the DC scheme, iterative scheme, and non-iterative scheme.
The left and right values in each cell denote the BP thresholds for the ID ($p_{\rm s}=0$) and IDS ($p_{\rm s}=0.04$) channels, where $\ast$ signifies $p_{\rm id}^{\ast} < 0.001$.}
\label{table:thr}
\scalebox{0.84}{
\begin{tabular}{|c||c|c|c|c|}
\hline
\diagbox{scheme}{code}& Bi-AWGN\cite{ldpc2}                    & ID\cite{my_irr}                  & IDS\cite{my_irr2}             & SC\cite{scid}                \\ \hline \hline
non-iterative& $\ast$, $\ast$ & $\ast$, $\ast$ & $\ast$, $\ast$        & $\ast$, $\ast$ \\  \hline
iterative& $\ast$, $\ast$ & \textbf{0.097}, $\ast$ & 0.072, \textbf{0.046}        & \textbf{0.097}, \textbf{0.048} \\  \hline
DC ($T_{\rm max}=3$)      & \textbf{0.072}, \textbf{0.018}                       & 0.024, $\ast$     & 0.048, $\ast$ & \textbf{0.072}, \textbf{0.017}                      \\  \hline
DC ($T_{\rm max}=7$)      & \textbf{0.088}, \textbf{0.035}                       & 0.038, $\ast$     & 0.063, 0.011            & \textbf{0.087}, \textbf{0.035}                      \\  \hline
DC ($T_{\rm max}=15$)     & \textbf{0.091}, \textbf{0.039}                       & 0.040, $\ast$     & 0.067, 0.018           & \textbf{0.090}, \textbf{0.040}                      \\  \hline
DC ($T_{\rm max}=31$)     & \textbf{0.091}, \textbf{0.041}               & 0.040, $\ast$     & 0.067, 0.019            & \textbf{0.090}, \textbf{0.041}                      \\ \hline \hline
SIR limit            & \multicolumn{4}{c|}{0.0997, 0.0499} \\ \hline
\end{tabular}
}
\end{table}
\subsection{Finite-length simulation}
\begin{figure}[t]
\centering
\includegraphics[width=.95\linewidth]{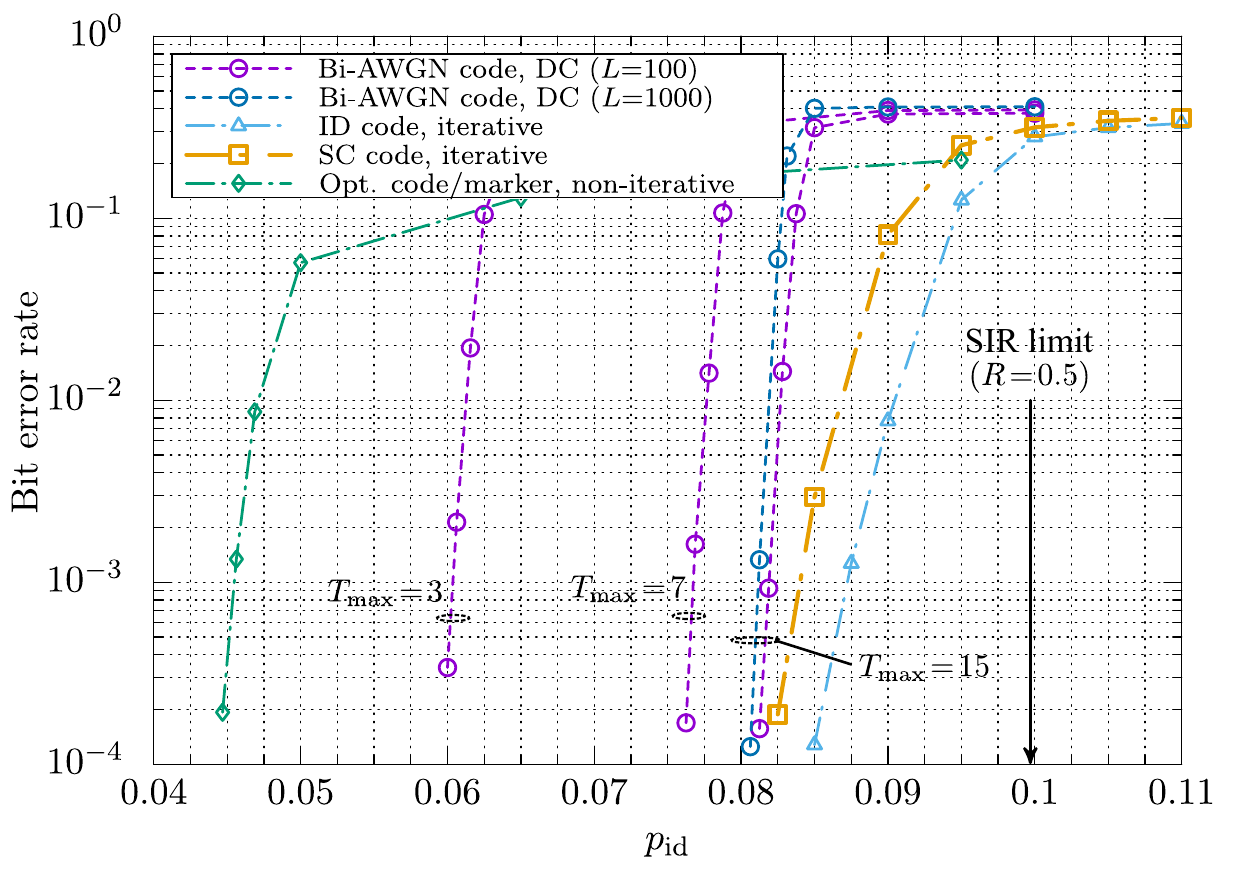}
\caption{BER curves of the proposed DC scheme with the Bi-AWGN code, iterative scheme with the ID and SC-LDPC codes, and non-iterative scheme with the concatenated LDPC/marker code in the ID channel ($p_{\rm s}=0$), where an LDPC code length of $n\approx 5\times 10^{4}$ is used.}
\label{fig:res3}
\end{figure}
\begin{table}[t]
\centering
\caption{Average number of executions of MAP detection for the DC and iterative schemes.}
\label{table:ave}
\scalebox{0.87}{
\begin{tabular}{|c||c|c|c|}
\hline
\multirow{2}{*}{\diagbox{$p_{\rm id}$}{code, scheme}} & Bi-AWGN\cite{ldpc2}& ID\cite{my_irr}& SC\cite{scid} \\ \cline{2-4}
                                       & DC ($T_{\rm max}=15$)                                      & \multicolumn{2}{c|}{iterative}                              \\ \hline \hline
0.070                                  & 16                                     & 21.3                                & 49.4                            \\ \hline
0.075                                  & 16                                     & 25.5                                & 57.9                            \\ \hline
0.080                                  & 16                                     & 31.7                                & 72.5 \\ \hline
\end{tabular}
}
\end{table}
In this section, we present the simulation results of finite-length LDPC codes with the DC scheme.
We compare the bit-error rate (BER) performance of the following schemes: (1) the DC scheme (we use $\bm{T}=(0,1,\ldots,T_{\rm max})$) with the Bi-AWGN code, (2) the iterative scheme with the $C(3,6,25)$ SC-LDPC and ID codes, and (3) the non-iterative scheme with concatenated rate-$0.5$ optimized LDPC and marker codes\cite{marker} (two-bit markers are inserted into an LDPC codeword every 10 bits).

Figure \ref{fig:res3} shows the BER curves with length $n\approx 5\times 10^{4}$ LDPC codes as a function of $p_{\rm id}$ with a sufficient number of iterations, where $p_{\rm s}=0$.
From the figure, the good performance can be obtained with the DC scheme for a wide range of $T_{\rm max}$, with a sharp waterfall drop in BER.
The BER curves of the DC scheme are positioned between the non-iterative and iterative schemes, and approach the latter when $T_{\rm max}=15$ at ID probabilities around $p_{\rm id}=0.08$.
Table \ref{table:ave} shows the average number of executions of MAP detection for the DC and iterative schemes.
Although it is clear from Fig.\ \ref{fig:res3} that the BERs of both schemes are vanishing (we did not observe any errors in our simulations), the DC scheme has smaller execution numbers compared with the iterative scheme.
At $p_{\rm id}=0.080$, in terms of the complexity of detection, the DC scheme is 2 and 4.5 times as efficient as the iterative scheme with the ID and $C(3,6,25)$ SC-LDPC codes, respectively.
\section{Conclusion}
This paper proposed the DC scheme for IDS channels, which employs delayed encoding and non-iterative chained detection and decoding strategies.
We presented the AIRs of the DC scheme over the channel and showed that for a large delay $T_{\rm max}$, the AIRs universally approach the SIRs without depending on substitution and ID occurrence probabilities.
We also analyzed the asymptotic performance of the DC scheme with LDPC codes, and showed that, with an appropriate LDPC code, the BP threshold approaches the corresponding AIR, even though the MAP detection is performed only once before BP decoding.
Moreover, the results of simulations conducted show that the DC scheme provides excellent performance with the finite-length codes while reducing complexity compared to the conventional scheme.
In future work, to approach the SIRs as closely as possible with a small delay and low complexity, we plan to design a delay scheme $\bm{T}$ and degree distribution of LDPC codes.
\section*{Acknowledgments}
This work was supported by JSPS KAKENHI Grant Numbers JP21K14160 and JP20K04473.

\end{document}